\documentclass[preprint]{aastex}
\usepackage{graphicx}
\begin{document}

\title{Planetary Torques as the Viscosity of Protoplanetary Disks}

\author{J. Goodman and R. R. Rafikov}
\affil{Princeton University Observatory, Princeton, NJ 08544}
\email{jeremy@astro.princeton.edu}

\begin{abstract}
We revisit the idea that density-wave wakes of planets drive accretion
in protostellar disks.  The effects of many small planets can be
represented as a viscosity if the wakes damp locally, but the
viscosity is proportional to the damping length.  Damping occurs
mainly by shocks even for earth-mass planets.  The excitation of the
wake follows from standard linear theory including the torque cutoff.
We use this as input to an approximate but quantitative nonlinear
theory based on Burger's equation for the subsequent propagation and
shock. Shock damping is indeed local but weakly so.  If all metals in
a minimum-mass solar nebula are invested in planets of a few earth
masses each, dimensionless viscosities [alpha] of order dex(-4) to dex(-3)
result.  We compare this with observational constraints.  Such small
planets would have escaped detection in radial-velocity surveys and
could be ubiquitous.  If so, then the similarity of the observed
lifetime of T Tauri disks to the theoretical timescale for assembling
a rocky planet may be fate rather than coincidence.
\end{abstract}


\section{Introduction}

The disks of young low-mass protostars (``protostellar disks'') are
observed to accrete at rates $\sim 10^{-8}M_\odot~\mbox{yr}^{-1}$
\citep*{HCGD}.
A comparable average accretion rate is implied by
the disk masses inferred from dust emission
($\sim 10^{-2\pm 1}~M_\odot$, \citealp{OB95}) when combined with the
maximum ages of T Tauri stars that show evidence for disks
($\sim 10^{6-7}~\rm{yr}$, \citealt{SSECS}), or 
with radiochemical estimates for the lifetime of the protostellar
nebula \citep{PC94}.  Indeed, the evidence that young protostars 
are surrounded by viscous accretion disks has only strengthened
in the quarter century since \citet{LP74} proposed it. 

While the effective viscosity of the hotter accretion
disks surrounding degenerate stars and black holes
is probably due to magnetohydrodynamic (MHD) turbulence
\citep[cf.][]{BH98}, protoplanetary disks may be
too cold, too weakly ionized, and too resistive to sustain MHD turbulence
\citep{BB94,Jin96,HS98} except perhaps at small radii,
in surface layers \citep{Gam96}, or during FU Orionis-type outbursts
\citep{HKH93}.  In view of several physical uncertainties
in the ionization balance \citep[cf.][]{Gam96}, and the possibly
subtle nature of the conductivity itself \citep{Wardle}, this conclusion
is only tentative.  But it is probably fair to say that the best
reason for supposing that the viscosity of protoplanetary disks
is magnetic remains the lack of a theoretically attractive alternative.
Convection \citep{LP80} and nonlinear shear
instability \citep{Zahn,Dubrulle} have been proposed, but these mechanisms 
appear to fail because of the very strongly stabilizing 
influence of a positive radial gradient in specific
angular momentum \citep{RG92,SB96,BHS}.

\citet{Larson89} suggested that density waves might provide the
efffective viscosity mechanism in protostellar disks.  As he
made clear, wave transport is rather different from viscous transport.
In the standard theory of thin accretion diks, viscosity ($\nu$)
plays two roles. Firstly, it provides an outward angular-momentum flux ($F_J$),
a torque exerted by the part of the disk interior to a given
radius on the part exterior.
Secondly, it heats the disk by dissipating the differential rotation.
In steady state, $F_J$ balances the inward flux of angular momentum
carried by the accreting matter, while the dissipation
balances the loss of its orbital energy.  Even if the disk is not steady,
provided only that it is truly viscous, then
\begin{displaymath}
F_J = - \nu\,\times\,2\pi\Sigma r^3\frac{d\Omega}{dr}
\end{displaymath}
where $\Sigma$ is the surface density and $\Omega$ the angular velocity.
Also, the dissipation rate per unit area is related to $F_J$ by
\begin{equation}\label{QJrel}
\dot E= -F_J\frac{d\Omega}{dr}.
\end{equation}
Any dissipative mechanism such that
$F_J\propto d\Omega/dr$ and eq.~(\ref{QJrel}) hold can be described
by an effective viscosity $\nu_{\rm eff}$; 
one expects this to be the case for dissipative mechanisms that are
sufficiently local.

Since density waves are not local, one may
deduce a different value of $\nu_{\rm eff}$ from $F_J$ than
from $\dot E$.  In particular, inviscid linear theory
\citep[henceforth GT80]{GT80} predicts that a planet embedded
in a gas disk excites 
density waves at the inner and outer Lindblad resonances of its
(circular) orbit; these waves carry $F_J>0$ but $\dot E=0$ by assumption,
so that $\nu_{\dot E}=0$ but $\nu_{J}>0$.
Actually, the theory implicitly assumes some damping
at large distance from the resonances, else the waves would reflect
from the edges of the disk and produce zero
net flux.  If one is concerned with the torques on the disk only,
then the damping length is unimportant as long as it is large.
But without dissipation, the waves cause
no secular changes to the orbits of the gas even at the
Lindblad resonances \citep{GN89}.

By demanding that eq.~(\ref{QJrel}) be satisfied,
\citet{Larson90} found that steady spiral shocks produce
\begin{equation}\label{alphaLarson}
\alpha\sim 0.013\left[(c/V)^3+0.08(c/V)^2\right]^{1/2}
\end{equation}
{\it independently of the source of excitation of the waves.}
Here $c$ is the sound speed and $V=r\Omega$ the orbital velocity.
For $c/V=0.03$, perhaps characteristic of the later phases of
the protosolar nebula at 1~AU, this gives $\alpha\approx 10^{-4}$. 
Larson's result depends upon a number of additional assumptions,
some inspired by earlier
work by \citet{Sp87} on self-similar shocks:
the waves were taken to form two spiral arms separated by $180^\circ$
in azimuth; and because they would be very tightly wrapped since $c\ll V$,
\citet{Larson90} approximated the wave profile by analyzing a
train of perfectly axisymmetric nonlinear waves.
Most importantly, at least by contrast with our own work reported below,
no source of excitation for the waves exists within the range of radii 
where eq.~(\ref{alphaLarson}) applies.  Larson argued, however, that
a bisymmetric spiral shock of the required strength could be launched
from the $2:1$ resonance of
a jovian-mass planet, and he deduced an accretion timescale
$\sim 10^{6-7}~\mbox{yr}$ in accord with observational requirements.

Jupiter-mass planets have been discovered during the past few years by
exquisitely sensitive radial-velocity surveys \citep{MCM}.  Most of
these planets have orbital periods shorter than that of Mercury.  They
are detected around $4-5\%$ of the stars surveyed, mostly G, F, and K
dwarfs.  There are indications that stars of higher-than-solar
metallicity are preferred \citep{Gon97,SIM,GLTR}.  Because of
observational selection effects, it is probably too early to derive
the incidence of giant planets in long-period orbits ($\sim 10~{\rm
yr}$).  If such planets are as rare as those on short-period orbits,
however, then we must seek another explanation for the observed
accretion rates and inferred lifetimes of most protostellar disks.
Nevertheless, it remains an intriguing notion that planet formation is
not just constrained by disk lifetimes, but may actually determine
them.

Our paper therefore inquires whether accretion could be driven by
planets of roughly terrestrial mass, which radial-velocity surveys
would not have detected.  Other things being equal, the torque exerted
on the disk by a planet is proportional to the square of the planet's
mass ($M_p$), so the influence of $M_p\sim 10^{-5}M_\odot$ might be
thought negligible compared with that of $M_p\sim 10^{-3} M_\odot$.
Just because its local effects are so strong, however, the larger
planet will almost certainly open a gap in the disk, whereas the
smaller one will not; since the torque per unit radius (smoothed over
resonances) decreases rapidly with distance from the planet (GT80),
the shorter range of its interaction may partly compensate for the
terrestrial planet's smaller mass.  At solar abundance there could be
tens of earth-mass planets in the disk, each too small to have
captured a significantly massive atmosphere.  We assume that the
planets are distributed through the disk in proportion to the local
surface density of the gas, neglecting orbital migration and radial
changes in the timescale of planet formation.  Unlike Larson,
therefore, we have a distributed source of excitation for density
waves, so that eq.~(\ref{alphaLarson}) need not apply (although
in fact we obtain a comparable value of $\alpha$ in the minimum-mass
solar nebula).

As discussed, no sensible effective viscosity can be derived for
density waves without considering their damping. \S2 shows for a
generic damping mechanism in a thin disk that if the waves are
absorbed over a distance small compared to the disk radius, but
possibly large compared to its thickness, and if the local damping
rate depends mainly on distance from the planet,
then eq.~(\ref{QJrel}) is satisfied so that a consistent value of
$\nu_{\rm eff}$ or $\alpha$ can in fact be defined.  However,
the viscosity is directly proportional to the damping length.
After briefly considering other mechanisms in \S3, we
conclude that shock damping is probably most important even at
$M_p\sim M_\oplus$.
In \S4, we calculate the excitation of the wake from linear theory, 
which is valid for sufficiently small $M_p$ [eq.~(\ref{Mlimit}].
In \S5 and the Appendix, we show that the subsequent nonlinear
propagation of the wake away from the planet can be reduced to
Burger's equation using some transformations and some
approximations that should be accurate for sufficiently low-mass
planets in sufficiently thin disks.
We describe quantitatively the formation of the shock and its subsequent
decay with increasing distance from the planet.  Most importantly,
we obtain formulae for the shock-damping length, and hence for
$\alpha_{\rm eff}$, as functions of $M_p$ and local disk parameters.
\S6 summarizes our main results and discusses their implications for
disks resembling a standard minimum-mass solar nebula.

\section{Effective $\alpha$-parameter.}\label{alpha_sect}

Let $f^{(0)}_J(M_p,r_p)$ be the angular-momentum flux excited by a
planet of small mass $M_p$ and orbital radius $r_p$, embedded
in a disk of local surface density $\Sigma$, sound speed $c$,
and angular velocity
$\Omega(r)\approx(GM_*/r^3)^{1/2}$, where $M_*$ is the mass of
the central star.
According to GT80,
\begin{equation}\label{pl_flow}
f^{(0)}_{J}(M_p,r_p)=(G M_p)^2 \frac{\Sigma r_p \Omega}{c^3}\times
\left\{ \frac{4}{9}\mu^3_{max}(Q)
\left[2K_0(2/3)+K_1(2/3)\right]^2\right\}.
\end{equation}
The dimensionless quantity $\mu$ is defined in such a way that
$\mu\Omega r/c\gg 1$ is roughly the largest azimuthal harmonic contributing
to the flux (the ``torque cutoff'');
it is a function of the gravitational stability parameter 
$Q=c\Omega/\pi G \Sigma$ \citep{To64}.
Selfgravity is probably very weak in the later evolutionary phases
of protostellar disks, so it is appropriate to take $Q\to\infty$.
In this limit, it follows from GT80's results that the quantity in
curly brackets $\approx 0.93$.  As acknowledged by GT80 and elaborated
by later authors, eq.~(\ref{pl_flow}) is valid only to leading order in
$c/\Omega r$, the ratio of disk thickness to radius.
At the next order,
there is a slight imbalance between the interior ($r<r_p$) and exterior
($r>r_p$) fluxes; the difference between the two torques represents
a net torque on the planet, whose orbit therefore
``migrates,'' probably towards smaller $r_p$
\citep[and references therein]{Wa97}.  The effective viscosity we
calculate depends on the sum of the interior and exterior fluxes,
so we neglect the difference between them.

The flux rises quickly to
a constant value (\ref{pl_flow}) over a radial distance from the planet
of order $l\equiv c/|rd\Omega/dr|^{-1}$, the length over which the
rotation velocity changes by Mach 1.  The strongest Lindblad resonances
giving rise to the wake lie within this distance.  Absent damping,
the angular momentum flux is radially constant at $|r-r_p|\gg l$.
We can characterize the strength of damping by the reciprocal of the
damping length $l_d$ at which $f_J$ falls to $\sim e^{-1}$ of
eq.~(\ref{pl_flow}).

The larger the damping length, the greater the dissipation of orbital
energy by the density-wave wake, for the following reason.
If a quantity $\Delta J$ of angular momentum is added to the disk in the
form of waves having angular pattern speed $\Omega_p$ (in steady state,
the wake has the pattern speed of the planet), the work required to
create the waves is $\Omega_p\Delta J$.  When these waves damp
at radius $r$, the work done on the mean flow there is $\Omega(r)\Delta J$.
The difference $[\Omega_p-\Omega(r)]\Delta J$ represents the energy
available to be dissipated, assuming that this is positive.  It can be shown
that $\Delta J$ is positive (negative) for waves excited exterior
(interior) to the planet in the approximation that
$\Omega_p=\Omega(r_p)$.
From these considerations, it follows that
the total mechanical energy dissipated by the wake of the one planet is
\begin{equation}\label{endot}
\dot e(M_p,r_p) = \int\limits_{0}^\infty \left[\Omega(r)-\Omega_p\right]
\frac{df_J}{dr}dr.
\end{equation}

Clearly $\dot e\propto l_d$, asssuming $l_d\ll r_p$.
Suppose that
the background properties of the disk are approximately constant over
radial distances of order $l_d$.
Write the angular momentum flux in terms
of its ideal asymptotic value (\ref{pl_flow}) and a dimensionless
distribution function, $f_J(r)= f_J^{(0)}\cdot\phi(r-r_p)$.
Our notation presumes a damping mechanism such that
$\phi$ depends much more rapidly on $r-r_p$
than on $(r+r_p)/2$, the latter dependence being significant
only on the scale of $r$ itself.  The integral (\ref{endot}) can
now be approximated by
\begin{equation}\label{dote}
\dot e(M_p,r_p) = f_J^{(0)}(M_p,r_p)\left(\frac{d\Omega}{dr}\right)_{r_p}
\int\limits_{-\infty}^\infty x\frac{d\phi}{dx}\,dx,
\end{equation}
where $x\equiv r-r_p$.
The lower limit of integration has been extended to $-\infty$ for convenience
since $\phi(x)\approx 0$  unless $|x|\ll r_p$.

Now suppose that there are many planets of the same mass distributed
through the disk in numbers $N(r)$ per unit radius, and that
$rN(r)\gg 1$.  Then the total flux at radius $r$ is (on average)
\begin{equation}\label{FJphi}
F_J(r) = \int\limits_0^{\infty} N(r_p)f_J^{(0)}(M_p,r_p)\phi(r-r_p)dr_p
~\approx~ N(r)f_J^{(0)}(M_p,r)\int\limits_{-\infty}^{\infty} \phi(x)dx,
\end{equation}
assuming that $N(r)$ and $f_J^{(0)}(M_p,r)$ vary only slowly with radius.
Similarly, the dissipation rate per unit radius at $r$ is
\begin{eqnarray}\label{Edotphi}
\dot E(r) &=& \int\limits_0^{\infty} [\Omega(r)-\Omega(r_p)]
N(r_p)f_J^{(0)}(M_p,r_p)\frac{\partial\phi}{\partial x}(r-r_p)\,dr_p
\nonumber\\
&\approx& \frac{d\Omega}{dr}\,N(r)f_J^{(0)}(M_p,r)
\int\limits_{-\infty}^{\infty} x\frac{d\phi}{dx}dx
~=~-\frac{d\Omega}{dr}\, F_J(r),
\end{eqnarray}
where the last step has used integration by parts.
So, as promised in \S1,
equation (\ref{QJrel}) is satisfied by density-wave
wakes under the assumptions of many planets and local damping.

In the standard theory of viscous accretion disks \citep{Pr81},
the viscous energy dissipation rate per unit radius is related to
the \cite{SS} viscosity parameter $\alpha$ by
(assuming $\Omega\propto r^{-3/2}$)
\begin{equation}\label{visflow}
\dot E_{\rm visc} = \alpha\,\frac{9\pi}{2}
r\Omega\Sigma c^2.
\end{equation}
Let the planets have 
average mass $\langle M_p \rangle$ and let them account for a fraction
$Z_p\ll 1$ of the total nebular mass;
we expect that $Z_p$ is comparable to the metallicity of
the gas if planet-formation is efficient but the planets are too small
to have captured massive gaseous envelopes \citep{MNH78,PHP93,INE00}
Then the average number of planets per unit radius is
\begin{equation}
N =\frac{2 \pi r Z_p \Sigma}{\langle M_p\rangle}.
\label{num_pl}
\end{equation} 
The total dissipation rate per unit radius due to planetary
wakes is $N\langle\dot e\rangle$, so by comparison with eq.~(\ref{visflow}),
\begin{equation}\label{def_alpha_eff}
\alpha_{\rm eff}= \frac{4}{9}\frac{Z_p}{\Omega c^2}
\frac{\langle\dot e\rangle}{\langle M_p\rangle}.
\end{equation}
Despite the relations (\ref{pl_flow}) and (\ref{dote}), we are not yet
in a position to evaluate $\alpha_{\rm eff}$ numerically because we
have not determined the width of the distribution $\phi$, or equivalently,
the damping length.

In the rest of the paper we assume 
a minimum mass solar nebula 
\citealp[MMSN, cf.][]{Ha81}:
\begin{eqnarray}\label{MMSNSig}
\Sigma(r)=1700~ {\rm g}~{\rm cm}^{-3} r_{AU}^{-3/2},\\
c(r)=1.2~ {\rm km}~{s}^{-1} r_{AU}^{-1/4},\label{MMSNc}
\end{eqnarray}
where $r_{AU}$ is the radius $r$ expressed in astronomical units.
For these parameters,
\begin{equation}
Q=67~r_{AU}^{-1/4},
\label{Q}
\end{equation}
which is effectively infinite for our purposes, so that
we completely neglect self-gravity of the disk.  The MMSN plausibly
corresponds to the state of protostellar disks as observed around
T Tauri stars, but disks are likely to have been more massive at
earlier evolutionary phases when the star was enshrouded by infalling
dust and gas.  Self-gravity could then have been important to the effective
viscosity and accretion rate \citep{Gam00}.

Another important assumption is that the sound speed is independent
of height above the midplane.
This is appropriate for the later evolutionary
phases of protostellar disks, when the local thermal structure is probably
dominated by passive reprocessing of radiation from the central star
\citep{CG97}, and it is in these phases that planets are most likely
to be dynamically important.
Vertical isothermality allows us to treat even short-wavelength
acoustic disturbances two-dimensionally, \emph{i.e.} in radius and
azimuth.  When the disk is prodominantly self-luminous, as 
is likely in earlier phases and during FU Orionis outbursts,
the temperature may decrease away from the midplane,
in which case acoustic waves will concentrate toward
the surface and perhaps damp more quickly than we estimate here
\citep{LO98,OL99}.  The thermal stratification of an active disk is
very uncertain, however, since the distribution of dissipation over
height is poorly understood; in some models, dissipation occurs mainly
near the surface, which would also tend to make the disk vertically
isothermal \citep{Gam96}.

\section{Linear damping}\label{linear}

Linear damping mechanisms are those for which the damping length
is independent of amplitude and which do not couple different
azimuthal harmonics of the wake.
One possibility is a background viscosity $\nu_0$
that acts on density waves as well as the mean shear.
\citet{TML} showed that in this case
the wave's angular momentum decays with distance as
(for definiteness, consider $r>r_p$)
$f_{J,m}\propto\exp\left[-2\int\limits^r \Delta k_\nu\,dr\right]$,
where $f_{J,m}$ is the angular momentum flux carried by the $m^{\rm th}$
azimuthal harmonic, and $\Delta k_\nu$ is the contribution to
the imaginary part of the WKBJ wavenumber due to viscosity:
\begin{equation}
{\rm Re}(k)^2\approx\frac{m^2(\Omega-\Omega_p)^2-\kappa^2}{c^2}.
\label{WKB}
\end{equation}
and
\begin{equation}
\Delta k_I\approx
\nu_0\left[\frac{4}{3}+\frac{\kappa^2}{m^2(\Omega-\Omega_p)^2}\right]
\frac{m(\Omega_p-\Omega)}{2c^2}{\rm Re}(k),
\label{imk}
\end{equation}
The most important contributions to the flux are for 
$m\sim \mu$, where $\mu\sim r/l\gg 1$ is the torque cutoff (GT80),
so the terms involving the epicyclic frequency $\kappa$ ($\approx \Omega$)
can be neglected beyond a few $l$ from the planet.
Hence $\Delta k_I\sim (r-r_p)^2/l^4$, so 
\begin{equation}
f_{J,m}\sim f_{J,m}^{(0)} \exp\left[-\alpha_0(r-r_p)^3/h^3\right].
\label{solut1}
\end{equation}
Here $\alpha_0\equiv \nu_0/\Omega h^2$ and $h\equiv c/\Omega r= (3/2)l$
is the disk thickness.
Hence the damping length is $ l_d\sim \alpha_0^{-1/3}h$.
If $\alpha_0$ is as large as $10^{-3}$, then there is no need
to invoke density waves to explain accretion.  
On the other hand,
if $\alpha_0\ll 10^{-3}$ then $l_d\gg 10 h$.

\citet{CW96} have proposed another linear damping mechanism: radiative
losses from the disk surface.  They conclude that the effect
is strong, but we believe that they made inappropriate use of the adiabatic
approximation.  \citeauthor{CW96}
estimate the radiative losses by evaluating the
adiabatic (lossless) density-wave eigenfunction at the disk
photosphere.  Because they take an isothermal background state, as
appropriate for a passive disk warmed by the central star, the
first-order temperature perturbation at the photosphere is as large as
it is at the disk midplane, so that their damping
rate is almost independent of the total optical depth of the disk
(cf. their Fig.~1).  This result is implausible.  By contrast, we find
that if the non-adiabatic effects are included self-consistently in
the linear analysis, then the damping rate is inversely proportional
to the total optical depth and is very much reduced compared to their
estimate.  The typical optical depths in the MMSN are
$\tau\sim 10^3-10^4$.

For the planetary wake, radiative
diffusion in the radial direction [neglected by \citeauthor{CW96}]
is much more important than losses from
the surface, because the radial
wavelength is typically shorter than the vertical thickness.
The process is similar to damping of a sound wave in air by thermal
conduction.  Adapting formulae from \citet{LL}, we find
\begin{equation}
\Delta k_I\sim\frac{k^2 h^2}{c}\frac{2 m_H\sigma_B T^3 }{\Sigma k_B \tau},
\end{equation}
where $\sigma_B$ is the Stephan-Boltzmann constant,
$2 m_H$ is the mass of the hydrogen molecule, and $k_B$ is
Boltzmann's constant (and $c$ is still the speed of sound, not of
light). Using eq.~(\ref{WKB}) for $k$, we find
that the damping length due to radiative diffusion is
\begin{equation}
 l_d\sim h\left(\frac{\Omega\Sigma k_B \tau}
{2 m_H \sigma_B T^3 }\right)^{1/3}.
\end{equation}
The most interesting feature of this expression is that now $ l_d\propto
\tau^{1/3}$, so that it is much smaller than in the case of radiative losses 
from surface where (in our analysis) $ l_d\propto\tau$.
Numerical estimates show, however, that the damping length is
still $\gtrsim r$ for typical conditions.

We conclude that the obvious linear damping mechanisms produce damping
lengths as large as the radius itself, or larger.
Therefore, these mechanisms are unimportant compared to shock damping which,
as will be seen, leads to a damping length that scales with
planetary mass as $\propto M_p^{-2/5}$
but is still only a modest multiple
of the disk thickness even if $M_p$ is as small as $M_\oplus$.

\section{Linear excitation of a planetary wake}\label{excite}

Even if the planet does not open a gap in the disk, there is a minimum
radial separation $|r-r_p|=l$ at which density waves can be excited
because at smaller distances the velocity of the gas in the corotating
frame of the planet is subsonic, and a stationary perturber does not
excite acoustic waves in a subsonic flow \citep{LL}.
This explains the torque cutoff (GT80), at least in nonselfgravitating
disk where the density waves are basically acoustic.
For a planet of sufficiently low mass [eq.~(\ref{Mlimit})], the forces
exerted at this minimum distance are weak enough so that the excitation
of the wake and its angular momentum flux can be accurately calculated
by linear theory.
It is well known that in the absence of linear damping (viscosity,
thermal conductivity), a sound wave of small but finite amplitude
propagating into still air eventually shocks;
the distance to the shock is proportional to the wavelength
and inversely proportional to the initial amplitude \citep{LL,Whitham}.
In a disk the shock length is shorter and the dependence on initial
initial amplitude is weaker because
differential rotation compresses the
radial width of the wake as it propagates.

Linear excitation has been studied by \citet[henceforth GT78]{GT78},
GT80, \citet{Artym}, \citet{Wa97}, and others.
The relative phases of the azimuthal Fourier harmonics of the wake
are normally discarded because they are not relevant to the
torque.
But the onset of the shock depends upon the local slope of the
wake front, which in turn depends upon these phases.  Therefore, 
we have repeated the linear calculation using the methods of GT78 and GT80. 

As usual, $x=r-r_p$ and $y=r_p(\theta-\theta_p)$, denote
pseudo-cartesian radial and azimuthal coordinates in a corotating
system centered on the planet, and all properties of the background
flow are expanded to lowest order in $x/r_p$.  The
background surface density $\Sigma$ and sound speed $c$ are made
constant, as are the rotation rate ($=$Coriolis parameter) $\Omega$,
shear rate $2A\equiv rd\Omega/dr$, and vorticity
$2B\equiv2(\Omega+A)$.  The background flow with respect to the planet
is $2Ax\mathbf{e}_y$.  No distinction is made between $\Omega$ and the
angular velocity of the planet, since this is not important for the
angular momentum flux carried by the wake to leading order in $h/r_p$
[but vital to the net torque on the planet (\citet{Wa97}].  Although
we are interested in keplerian disks, for which $2A/\Omega=-3/2$ and
$2B/\Omega=+1/3$, the physics is clarified by retaining $A$ and $B$ in
basic formulae.  We use the Mach-1 distance $l\equiv c/|2A|$ rather
than $h=c/\Omega=(3/2)l$ as a reference length.  A convenient unit for the
planetary mass is
\begin{equation}\label{Munit}
M_1\equiv \frac{c^3}{|2A|G},
\end{equation}
which reduces to $2c^3/3\Omega G$ in a keplerian disk. 
In linear theory, the amplitude of the wake is $\propto M_p$, so
it is sufficient to calculate as if $M_p=M_1$ and
scale the results accordingly.  Note that $M_p\gtrsim M_1$ is also the
point at which the linear approximation begins to fail; the
Roche lobe of the planet becomes $\gtrsim l$, and a gap may open
in the disk (\S6).

The wake is steady in the planet frame, so that
its perturbations in radial and azimuthal velocity,
$u,v\ll c$, and the relative perturbation in surface density,
$\sigma=\delta\Sigma/\Sigma\ll 1$,
are functions of $(x,y)$ only.   Their spatial Fourier transforms
$\hat u$, $\hat v$, and $\hat \sigma$ $(k_x,k_y)$ satisfy 
(GT78, GT80)
\begin{eqnarray}\label{linveqn}
\frac{d^2\hat v}{d\tau^2} &+& [c^2k(\tau)^2 + \kappa^2]\hat v
= -ik_y\frac{d\hat\phi}{d\tau}+2ik_x(\tau)B\hat\phi,\nonumber\\
\hat u&=& -\frac{1}{c^2k_y^2+4B^2}\left(2B\frac{d\hat v}{d\tau}
-c^2k_xk_y\hat v+2Bik_y\hat\phi\right),\nonumber\\
\hat \sigma&=& \frac{i}{c^2k_y^2+4B^2}\left(k_y\frac{d\hat v}{d\tau}
+2Bk_x\hat v+ik_y^2\hat\phi\right).
\end{eqnarray}
The timelike variable $\tau$  is related to the pitch angle of the wave:
\begin{displaymath}
\tau\equiv -\frac{k_x}{2Ak_y}\,;
\end{displaymath}
the notation is standard and will not be confused with optical
depth.
An individual Fourier component evolves at constant $k_y$ but varying $k_x$
due to the differential rotation.
Since $A<0$, all Fourier information flows along the
characteristics $dk_x/d\tau=-2Ak_y$ from leading ($\tau<0$) to trailing
($\tau>0$) waves, and eqs.~(\ref{linveqn}) describe the evolution of
the amplitudes along those characteristics.
The quantity $k$ is the instantaneous wavenumber $\sqrt{k_x^2+k_y^2}$, and
$\hat\phi= -2\pi GM_p/k$ is 
the Fourier transform of the point-mass planetary potential.
Appropriate initial conditions are $\hat v=0$ at $\tau=-\infty$.
Unlike GT80, we have neglected the selfgravity of the disk.

Unless otherwise noted, the rest of this section assumes units
$c=-2A=1$, so that $l=1$, and the wake amplitude is scaled to $M_p=M_1$.

We solved eqs.~(\ref{linveqn}) numerically to obtain $(\hat u,\hat
v,\hat\sigma)$ on a grid of $N_x\times N_y=4096\times 8192$ points in
$(k_x,k_y)$.  The very large grid proved
necessary to minimize numerical artifacts of the discrete
Fourier transform (DFT).  We chose the maximum (Nyquist) value of $k_y=8$,
well beyond the torque cutoff.  The
corresponding azimuthal grid resolution and periodicity length
are $\Delta y=(\pi/8)$ and $N_y\Delta y\approx 3217$,
respectively.  Since the locus of the wake is approximately
$y\approx -x^2\mbox{sign}(x)/2$, we chose
$L_x= 2\sqrt{L_y}\approx 113.$, implying a radial
resolution $\Delta x=L_x/N_x=0.028$.  

The Fourier data were filtered in pitch angle to avoid aliasing.
When $k_x\gg k_y~\&~\kappa$, the
homogeneous solutions to eqs.~(\ref{linveqn}) change phase by
$k_x\Delta\tau=\tau\Delta k_x$ from $k_x$ to $k_x+\Delta k_x$, where $\Delta
k_x=2\pi/L_x$ is the resolution in Fourier space, and $k_x=\tau k_y$
in the present units.
This phase change should be $<\pi$, in other words 
$\tau_{\max}= L_x/2\approx 57$, else the phase will be aliased and the
inverse DFT will assign the wave to the wrong $x$ position.
We multiplied the Fourier components by the pitch-angle filter
\begin{displaymath}
W(k_x,k_y)=\cases{1 & if $\tau<\tau_{\max}/2$, \cr
              2(1 -\tau/\tau_{\max})& $\tau_{\max}/2\le \tau<\tau_{\max}$,\cr
                  0 & $\tau\ge \tau_{\max}$.}
\end{displaymath}
Because the pitch angle increases linearly with $|x|$
in coordinate space, this filter attenuates the wake at $|x|\gtrsim L_x/4$.
We also used a similar filter in the $k_y$ direction, tapering linearly
from unity at $|k_y|=k_{y,\rm Nyquist}/2$ to zero at
$|k_y|=k_{y,\rm Nyquist}$.

We checked our calculation against the angular momentum flux reported
by GT80 in two different ways.
One can show that as $L_x,L_y\to\infty$,
\begin{equation}\label{Fflux}
f_J = \lim_{\tau\to\infty}~\frac{ r_p\Sigma}{4\pi^2 l}
\int\limits_{0}^\infty \langle\hat u(k_x,k_y)\hat v^*(k_x,k_y)
\rangle_{\tau}\,|k_y|\,dk_y,
\end{equation}
the average being taken over oscillations in $\tau$ at fixed $k_y$.
We evaluated a numerical approximation to this formula in Fourier space
(before applying the windows discussed above)
and found agreement with the flux (\ref{pl_flow}) reported by GT80
for $Q=\infty$ to the accuracy they quoted, \emph{viz.} two significant
digits.  We also evaluated the flux as a function of $x$ in coordinate
space using eq.~({\ref{GTflux}).  The errors are larger because
the coordinate-space wake is softened by the pitch-angle and $k_y$ filters,
but nevertheless $f_J(x)$ is nearly constant and correct
within $\pm 3\%$ between $|x|=6$ and $|x|= 28$.

Fig.~\ref{Fig1} shows that beyond a few Mach lengths ($l$)
from the planet, the linearized wake profile is approximately constant
as a function of $y~+~x^2\mbox{sign}(x)/2$, as expected
if the source terms are negligible at $|x|\gg 1$ and the wake
propagates as a tightly-wrapped acoustic disturbance (see below).
To the extent that this is true, the profiles can be regarded
equally well as azimuthal cuts across the wake
at fixed radius, or as radial cuts at fixed azimuth but
with $x^2/2$ rather than $x$ as radial coordinate.
If plotted against $x$, the profile would shrink
in width $\propto |x|^{-1}$ for $|x|\gg 1$.
The profile is about as compact as possible given the torque cutoff
[the integrand of eq.~(\ref{Fflux}) peaks at $k_y\approx 0.36\,l^{-1}$],
showing that the phases of the Fourier components of the wake are
closely aligned.
\begin{figure}
\includegraphics[width=5in]{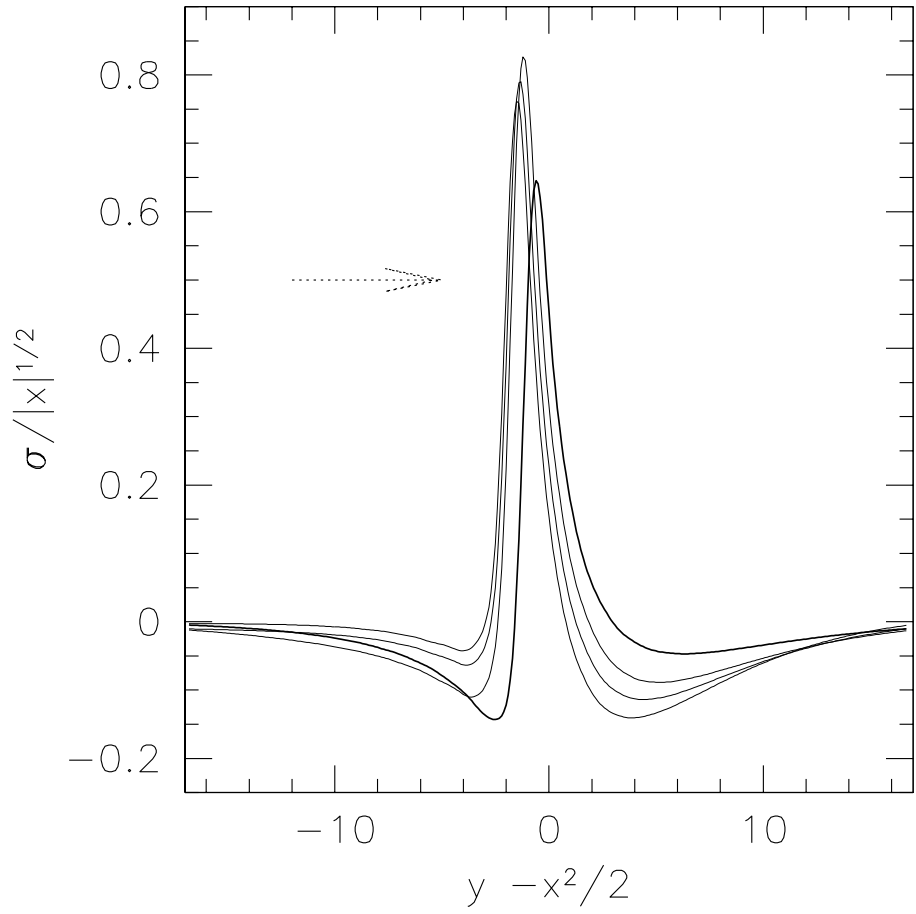}
\caption{Surface-density profile of the linearized wake at
$x=-2$ (heavy line), $-4,-6,-8$.
Division by $|x|^{1/2}$ removes growth due to flux conservation.
Arrow shows direction of gas flow relative to profile.
See text for dimensional units.
\label{Fig1}}
\end{figure}

\section{Nonlinear propagation}

The fully nonlinear steady-state fluid equations in $xy$ are
\begin{eqnarray}\label{nonlin0}
(2Ax+v)\partial_y u + u\partial_x u -2\Omega v+ \frac{c^2}{\Sigma}
\partial_x\Sigma &=& 0,\nonumber\\
(2Ax+v)\partial_y v + u\partial_x v +2B u+ \frac{c^2}{\Sigma}
\partial_y\Sigma &=& 0,\nonumber\\
(2Ax+v)\partial_y\Sigma + u\partial_x\Sigma +\Sigma\partial_x u+\Sigma
\partial_yv &=& 0.
\end{eqnarray}
Henceforth $\Sigma$ denotes the full surface density, and
$\Sigma_0$ its unperturbed value.
The sound speed follows the adiabatic law
$c^2(\Sigma)= c_0^2(\Sigma/\Sigma_0)^{\gamma-1}$,
a good approximation for weak shocks.
We have dropped the planetary source terms, since at distances
$\gg l$ they are unimportant.

The system (\ref{nonlin0}) is hyperbolic, and 
in the Appendix, we demonstrate that for $|x|\gg l$,
it can be reduced  
to a single first-order nonlinear equation:
\begin{equation}\label{chieqn}
\partial_{t}\chi-\chi\partial_\eta\chi = 0,
\end{equation}
which is the inviscid Burger's equation \citep{Whitham}.
The dimensionless variables appearing here are related to radius, azimuth,
and density contrast as follows:
\begin{eqnarray}
{t} &\equiv& \frac{4}{5}\left|\frac{y}{l}\right|^{5/4}
~\approx~ \frac{2^{3/4}}{5}\left|\frac{x}{l}\right|^{5/2},\label{taudef}\\
\eta &\equiv& \frac{y}{l}+\frac{x^2}{2\,l^2}\mbox{sign}(x),\label{etadef}\\
\chi &\equiv& \left|\frac{y}{l}\right|^{-1/4}
\frac{\gamma+1}{\gamma-1}\cdot
\frac{c-c_0}{c_0}~\approx~
\left|\frac{y}{l}\right|^{-1/4}
\frac{\gamma+1}{2}\cdot
\frac{\Sigma-\Sigma_0}{\Sigma_0}~.\label{chidef} 
\end{eqnarray}

\subsection{The shock}

For almost any smooth choice of initial conditions in Burger's equation
(\ref{chieqn}),
the solution will eventually threaten to become 
double-valued, so that one must fit in a shock \citep{Whitham}.
The identity
$(\partial_{t}\chi)_\eta/(\partial_\eta\chi)_{t}=-(\partial_{t}\eta)_\chi$
converts eq.~(\ref{chieqn}) to the linear equation
\begin{equation}\label{linB}
\left(\frac{\partial\eta}{\partial {t}}\right)_\chi= -\chi,
\end{equation}
which says that each point on the wave profile moves forward at a
``speed'' determined by the value of $\chi$ at that point.  A shock
develops when higher (larger $\chi$) parts of the profile overtake lower
ones.
The first shock develops where the initial profile is steepest, 
after a delay
\begin{equation}\label{taushock}
{t}_{\rm shock}-{t}_0= \left[\max\left( 
\frac{\partial\chi}{\partial\eta}\mbox{sign}\chi\right)
\right]_{0}^{-1}.
\end{equation}
If the initial profile changes sign, as it does in Fig.~(\ref{Fig1}),
there will be at least two shocks eventually: one moving to the
left and one to the right in this reference frame, which is moving
radially away from the planet at $c_0$.

\begin{figure}
\includegraphics[width=5in]{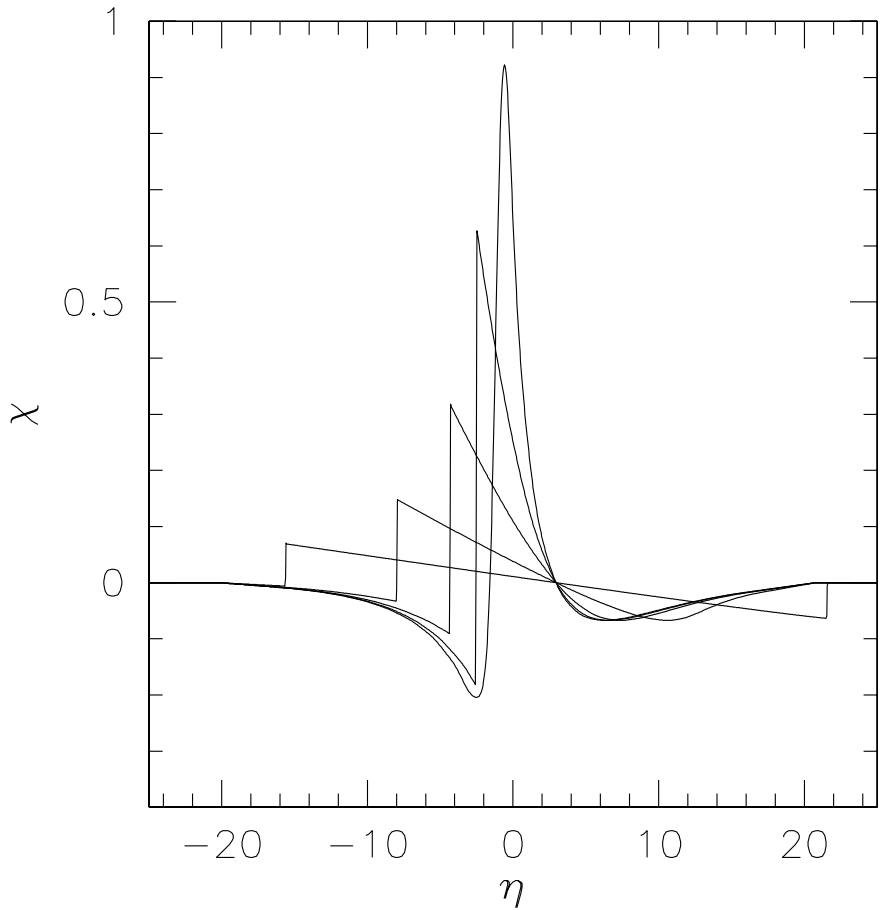}
\caption{Nonlinear evolution of the wake profile according to
eqs.~(\ref{chieqn})-(\ref{chidef}), integrated \emph{via}
second-order advection scheme \citep{vanL}.  Initial conditions 
taken from linear wake profile at $x_0=-2\,l$ (${t}=1.903$)
scaled to planet mass $M_p=M_1$.
Profiles shown at ${t}-{t}_0=0,4,16,64,256$ ({\it highest to lowest}).
The latest profile has the expected N-wave shape.
\label{Fig2}}
\end{figure}

The shock jump condition depends upon what is to be conserved.
In our case, $\chi$ is approximately proportional to
the density contrast $(\Sigma-\Sigma_0)/\Sigma_0$ provided that the latter
is small, so mass conservation demands
\begin{equation}\label{jump}
(\chi_1-U_s)\chi_1 = (\chi_2-U_s)\chi_2\quad\Leftrightarrow\quad
U_s=\frac{\chi_1+\chi_2}{2},
\end{equation}
where, $\chi_1$ and $\chi_2$ are the values of $\chi$ before 
and after the shock, and $U_s\equiv (d\eta/d{t})_{\rm shock}$ 
is the ``velocity'' of the shock
in these variables.  In physical variables, the radial velocity of the
shock is the mean of the sound velocities on either side of the jump,
as usual for weak shocks.

For an initial  pulse that changes
sign and has finite range, the final asymptotic behavior
is an ``N-wave,'' called such because the profile resembles that
letter of the alphabet (Fig.~\ref{Fig2}).  The unperturbed gas
flows from left to right in these variables, so it first encounters
the lefthand shock, followed by a rarefaction wave in which the
pressure falls below its unperturbed value, and finally
a jump back to the unperturbed conditions at the righthand shock.
The point where $\chi=0$ (on the rarefaction wave) does not move,
and the flux vanishes there, so mass conservation implies that
the areas in the left and right lobes of the N-wave
are both constant.
From this and the jump condition (\ref{jump}), it follows that
the height and width of the N-wave scale as ${t}^{-1/2}$ and 
${t}^{+1/2}$, respectively \citep[cf.][]{Whitham,LL}.

We choose $x_0=-2\,l$ as the matching point between the linear and
nonlinear calculations.  This is not far enough from the planet to
be completely outside the excitation region: the
flux in the linear wake is about $7\%$ smaller than its asymptotic value.
For $M_p\ll M_1$ one should probably use a larger value of $|x_0|$.
In practice, the interesting values of $M_p$ will be not much smaller
than $M_1$, and it is then necessary to match at a rather small $|x_0|$
in order to remain within the range where the linear calculations are valid.

Applying the criterion (\ref{taushock}) to the linear wake profile
at $|x_0|=2\,l$ [see Fig.~(\ref{Fig1})], whose amplitude is $\propto M_p$,
we find that the shock develops at 
${t}_{\rm shock}-{t}_0\approx 0.79(M_1/M_p)$.
The constant ${t}_0\approx 1.89$ is $\propto |x_0/l|^{5/2}$
but independent of $M_p$.  In order that subsequent formulae have a simple
power-law scaling with $M_p$, we shall neglect ${t}_0$, but it becomes
important as $M_p$ approaches $M_1$.
Using the relations (\ref{taudef}) and $h\equiv c/\Omega=(2/3)l$, 
${t}_{\rm shock}$ translates to
\begin{equation}\label{xshock}
|x|_{\rm shock}\approx 0.93
\left(\frac{\gamma+1}{12/5}\,\frac{M_p}{M_1}\right)^{-2/5}~h.
\end{equation}
This result indicates that if $M_p\ge M_1$, then the shock begins
immediately and cannot be separated from the excitation region.
Note that for $M_*=M_\odot$,
\begin{equation}\label{Mlimit}
M_1= \frac{2\,c^3}{3\,\Omega G}\approx 8.0
\left(\frac{c}{1~{\rm km~s^{-1}}}\right)^3
\left(\frac{r_p}{1~{\rm AU}}\right)^{3/4}~
 M_\oplus.
\end{equation}

\subsection{Angular momentum flux and energy dissipation}

By summing over the contribution of each azimuthal harmonic as given
by GT80, one can show that the angular-momentum flux carried
by the wake in a nonselfgravitating disk is
\begin{equation}\label{GTflux}
f_J(x)= \frac{c_0^3\,r_p}{|2Ax|\Sigma_0}\int\limits_{-\pi/r_p}^{\pi/r_p}
(\Sigma-\Sigma_0)^2\,dy.
\end{equation}
In terms of our dimensionless variables,
\begin{eqnarray}\label{action}
f_J(x)&=& \frac{2^{3/2}r_p\, c_0^3\Sigma_0}{(\gamma+1)^2\,|2A|}
~\Phi({t}),\nonumber\\[10pt]
\mbox{where}\quad\Phi({t})&\equiv& \int \chi^2(\eta,{t})\,d\eta.
\end{eqnarray}
We have used the fact that the wake depends
upon radius and azimuth most rapidly in the combination $\eta$,
and that its slowly-varying amplitude can be thought of as a function of
radius or azimuth according to the correspondence
$|x|\approx\sqrt{2l|y|}= 2^{1/2}(5{t}/4)^{2/5}l$.
Without shocks, the dimensionless flux
$\Phi$ is strictly conserved; in fact, the integral over $\eta$
of \emph{any} function of $\chi$ is conserved.

Once the shock develops, $f_J$ decays with distance as the angular
momentum carried by the wake is transferred to the mean flow.
From the scalings for the N-wave discussed above, it
follows that $\Phi({t})\propto {t}^{-1/2}$ at large ${t}$, hence
\begin{equation}
f_J(x)\propto |x|^{-5/4}\quad(|x|\gg |x|_{\rm shock}),\label{FJ}
\end{equation}
which was also confirmed by direct analytical N-wave expansion
of the original set of equations (\ref{nonlin0}).
The full evolution of $\Phi({t})$ is shown in Fig.~\ref{Fig3}.

The total energy dissipation rate associated with the absorption
of the wake is [cf. eq.~(\ref{endot})]
\begin{eqnarray*}
\dot e&=& 2\int\limits_{0}^\infty \left[\Omega(r)-\Omega(r_p)\right]
\frac{df_J}{dx}\\[10pt]
&\approx& \left(\frac{4}{5}\right)^{3/5}\left(\frac{2}{\gamma+1}\right)^2
c^3\,l\,\Sigma\int\limits_0^\infty \Phi({t})\,{t}^{-3/5}\,d{t},
\end{eqnarray*}
where we have used integration by parts to transfer the radial
derivative from $f_J$ to $\Omega(r)$.
Since ${t}^{-3/5}\Phi({t})\propto{t}^{-11/10}$
at large ${t}$,  the integral converges, but slowly; its value is
$\approx 12.3$ when $M_p=M_1$.

By scaling our numerical results to other $M_p$ and $\gamma$,
we find that the dissipation produced by a small planet is
\begin{equation}\label{shockdisp}
\dot e \approx
11. \left(\frac{\gamma+1}{7/5+1}\right)^{-2/5} c^3\,l\,\Sigma
\left(\frac{3\Omega GM_p}{2 c^3}\right)^{8/5}
\end{equation}
in a keplerian disk.
\begin{figure}
\includegraphics[width=5in]{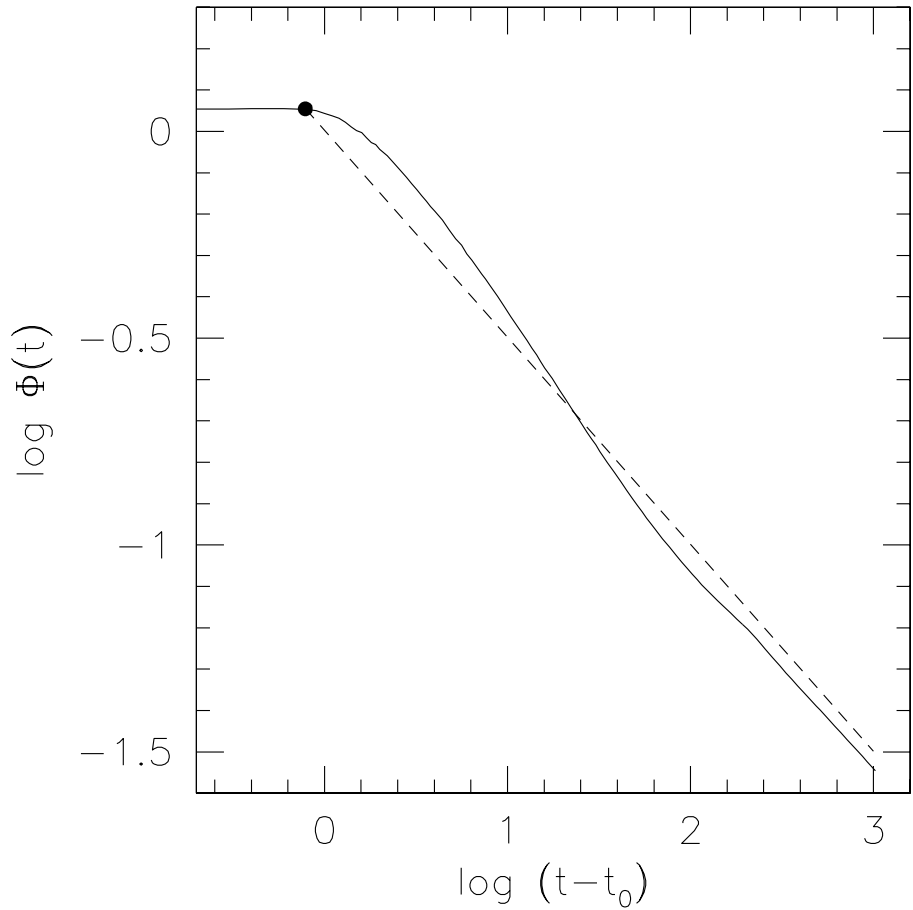}
\caption{{\it Solid line}: dimensionless angular momentum flux
[eq.~(\ref{action})] carried by the wake
versus propagation pseudotime (\ref{taudef}), based on numerical integrations
shown in Fig.~\ref{Fig2}.
{\it Dashed}:  Asymptotic ${t}^{-1/2}$ scaling appropriate to the N-wave,
normalized to agree with the constant preshock value $\Phi({t}_0)$
at ${t}={t}_{\rm shock}$ (\emph{bullet}).
\label{Fig3}}
\end{figure}
From eq.~(\ref{def_alpha_eff}), it now follows that
effective viscosity parameter is
\begin{eqnarray}
\alpha_{\rm eff}
&=&3.2~\left(\frac{\gamma+1}{7/5+1}\right)^{-2/5}
\frac{r^3 c^2 Z_p \Sigma}{G M_\star\langle M_p\rangle}
\left\langle\left(\frac{3\Omega G M_p}{2 c^3}\right)^{8/5}\right\rangle
\nonumber\\
&\approx& 1.2\times10^{-4}\left(\frac{r}{{\rm AU}}\right)^{-1/5}
\left(\frac{Z_p}{0.01}\right)
\left(\frac{M_p}{M_\oplus}\right)^{3/5}.
\label{alpha_eff}
\end{eqnarray}
For simplicity we have taken all planets to have the same mass in
the final line above, and we have used the MMSN model
(\ref{MMSNSig})-(\ref{MMSNc}).

In the Abstract, we stated that shock damping is local but only weakly
so.  It is local in the sense that most of the contribution to $\dot
e$ would come from a finite range of $|x|/h$ in the limit of a very
thin disk, $r/h\to\infty$.  But quantitatively, since protostellar
disks are not extremely thin, the results of this section show that 
the dissipation will extend over a range
comparable to the radius.  Although $f_J(x)$ falls to half of its full
value (\ref{pl_flow}) at $|x|\approx 2.1~|x|_{\rm shock}$, half of the
dissipation occurs beyond $9.1~|x|_{\rm shock}$, and the residual
contribution from distances $>|x|$ falls off as $|x|^{-1/4}$ when
$|x|\gg |x|_{\rm shock}$.

\section{Discussion}\label{disc}

We have shown that density-wave wakes excited by planets in a
nonselfgravitating protostellar disk serve as an effective
viscosity mechanism, provided that the waves are damped over a
distance smaller than the orbital radius of the planet.  The
effective viscosity is proportional to the damping length, which is
most likely to be limited by shocks unless the mass per planet $M_p\ll
M_\oplus$.  If a fixed total mass is available to form planets, then
$\alpha_{\rm eff}\propto M_p^{3/5}$.  
Eq.~(\ref{alpha_eff}) suggests
that $\alpha_{\rm eff}\sim 10^{-4}$ in the inner solar system where
$M_p\sim M_\oplus$, and $\alpha_{\rm eff}\sim 3\times 10^{-4}$ at
$r\sim 10~AU$ if $M_p\sim 10~M_\oplus$, which is
characteristic of the rock+ice \emph{cores} of the giant planets,
though very much less than the total mass of Jupiter or Saturn
\citep{Gu99}.

Our estimates for $\alpha_{\rm eff}$ may seem
optimistic because they assume that all of the rocks and ices are
efficiently converted into planets.
On the other hand, $\alpha_{\rm eff}$ could be even larger if $M_p$ is
augmented by gas.  In that case, $\alpha_{\rm eff}\propto M_p^{8/3}$ 
because the mass in planets is not limited to the metals.
There are two restrictions on the amount of gas we can allow for.
The first is technical: we have calculated the excitation of the density
waves from linear theory, which cannot be
justified when $M_p$ exceeds the limit (\ref{Mlimit}). This could be
overcome by nonlinear two-dimensional numerical calculations
\citep[\emph{e.g.~}][]{KP96}.
The other restriction is more substantial: once $M_p$ is large enough
to open a gap in the disk, the dissipation rate
probably scales more slowly than $M_p^{8/3}$.

Several estimates exist for the minimum mass required to open a gap,
$M_{\rm gap}$.  \citet{LP93} and \citet{BCLNP} argue that in a
completely inviscid disk, $M_{\rm gap}\approx M_1$ as given by
eq.~(\ref{Mlimit}) because of radial gas pressure gradients; recall that
at this mass, the Roche lobe (or Hill sphere) of the planet is $\approx h$.
\citet{Wa97} argues for $M_{\rm gap}\approx 0.14\pi\Sigma h r$, 
or $M_{\rm gap}\approx 0.8 (r/AU)^{3/4}\,M_\oplus$
in the fiducial MMSN (\ref{MMSNSig})-(\ref{MMSNc}).  This results from
comparing the time to open a gap with the time to migrate across it,
assuming that planetary wake damps immediately at the resonances where
it is excited.  Presumably $M_{\rm gap}$ should be increased to
allow for the finite damping length that we have
computed, but it would still be proportional to $\Sigma$, unlike
eq.~(\ref{Mlimit}).

$M_{\rm gap}$ is expected to be larger in a viscous disk than in an
inviscid one because the torque exerted by the planet can be balanced
by a viscous torque.
\citet{BCLNP} cite
\begin{displaymath}
\frac{M_{\rm gap}}{M_*}=\frac{40\nu}{\Omega r^2}
\approx 40\alpha\left(\frac{h}{r}\right)^2,
\end{displaymath}
which is again independent of $\Sigma$ (if $\alpha$ is).
In the MMSN, this becomes
\begin{displaymath}
M_{\rm gap}\approx 2.\left(\frac{\alpha}{10^{-4}}\right)
\left(\frac{r}{AU}\right)^{1/2}~M_\oplus.
\end{displaymath}
In our case, it seems likely that the shocks created by one planet
will resist the opening of a gap next to an adjacent planet, so it
is perhaps reasonable to use the value (\ref{alpha_eff}) for $\alpha$
on the righthand side above.

In the discussion to follow, we shall follow
\citet{LP93} and adopt (\ref{Mlimit}) as a provisional value
for $M_{\rm gap}$, but clearly the question deserves further study.
Then $M_{\rm gap}\sim 45~M_\oplus$ at $10~{\rm AU}$ using
eq.~(\ref{MMSNc}).  If the planets at
$10~{\rm AU}$ accrete enough gas to reach this limit (Saturn exceeds
it by a factor $\sim 2$), then $\alpha_{\rm eff}\approx 3\times 10^{-3}$.

In short, values for $\alpha_{\rm eff}$ in the range from $10^{-4}$ up
to at least $10^{-3}$ seem plausible for the density-wave mechanism
in a minimum-mass solar nebula.  The predicted accretion timescale is
\begin{displaymath}
t_{\rm acc}= \alpha^{-1}\left(\frac{r}{h}\right)^2\Omega^{-1}
\approx 5\times 10^6 \left(\frac{\alpha}{10^{-3}}\right)^{-1}
\left(\frac{r}{30~{\rm AU}}\right)
~\mbox{yr},
\end{displaymath}
in reasonable agreement with the disk lifetimes cited in \S 1.

By comparing observed sizes, ages, and accretion rates of T Tauri
disks with self-similar models, \cite{HCGD} estimate $\alpha\sim
10^{-2}$, which may be too large to be explained by planetary wakes,
especially if one must rely on the rocky core masses only.  The
strongest lower bounds on $\alpha$ assume that the disks started out
much more compact than observed.  In view of the wide scatter and
uncertainties in the observationally derived parameters, the
questionable relevance of self-similar models, and the theoretical
uncertainties just discussed, it is unclear whether \citeauthor{HCGD}'s
result is a major difficulty for us. Another concern
is that planets are likely to migrate inwards with respect to the gas
\citep{Wa97}, perhaps enhancing the viscosity of the inner disk but
diminishing that of the outer parts where most of the gas probably
resides.  Still another is that our mechanism operates only after
planet growth, which standard models place late in the lifetime of the
gaseous nebula, if not afterwards \citep{PHP93,INE00}, whereas
\citeauthor{HCGD} find \emph{larger} accretion rates in younger disks
on average.

All of these objections have merit.  Nevertheless, the density-wave
mechanism is interesting from several points of view.  First, it
provides a minimum effective viscosity when other mechanisms are in
abeyance.  For example, MHD turbulence may dominate only during FU
Orionis-like outbursts when the temperature and ionization of the disk
rises \citep{Gam98}; self-gravity may dominate in young, massive, cool
disks \citep{LP87}; but neither is likely to be effective in the late,
thermally passive phases when it is believed that planets form.
Second, calculable models for the effective viscosity are so difficult
to come by that it is worthwhile to examine any plausible candidate,
even if others seem more likely to dominate (but cannot be
calculated).  Third, the formalism offered in \S 5 and the Appendix
for the development and damping of the shock, though only approximate,
is expected to be reasonably accurate for weak shocks and simple
enough to use that it may find other applications.\footnote{Burger's
equation can be solved analytically given
a closed form for the initial wave profile \citep[cf.][]{Whitham}.}
Finally, there may be other nonaxisymmetric structures in protostellar
disks of greater mass than planets, for example vortices
\citep{AW95,GL00,Na00}.  Any such structure would create a
density-wave wake.

\acknowledgements We thank Charles Gammie, Peter Goldreich, Brad
Hansen, Kristen Menou, and Scott Tremaine for helpful discussions.
This work was supported by the NASA Origins Program under grant
NAG5-8385.

\appendix
\section{Reduction to Burger's equation}

At $|x|\gg l$, eqs.~(\ref{nonlin0}) are hyperbolic, as usual for
supersonic steady flows \citep{LL}: that is, they describe
an initial-value problem.
It is convenient to regard $y$ as the timelike variable.
The three characteristic ``velocities'' are
\begin{displaymath}
\left(\frac{dx}{dy}\right)_{0,+,-}=\frac{u}{2Ax+v},~
\frac{u(2Ax+v)\pm c\sqrt{(2Ax+v)^2+u^2-c^2}}{(2Ax+v)^2-c^2}.
\end{displaymath}
Eqs.~(\ref{nonlin0}) simplify for $|x|\gg l$, where
a linearized WKB treatment suggests that
$\partial_y\ll\partial_x$ and $v\ll u$.
Dropping $v$ entirely, and $\partial_y$ except in the combination
$2Ax\partial_y$, results in
\begin{eqnarray}\label{nonlin1}
\partial_y u + u\partial_{\xi} u 
+ \frac{c^2(\Sigma)}{\Sigma}\,\partial_{\xi}\Sigma &=&0\nonumber\\
\partial_y \Sigma + u\partial_{\xi} \Sigma + \Sigma\partial_{\xi} u &=&0.
\end{eqnarray}
We have replaced $x$ with $\xi\equiv Ax^2$ so that the equations are
autonomous, \emph{i.e.}, the independent variables do not appear explicitly.

The system (\ref{nonlin1}) is formally identical to one-dimensional
isentropic gas dynamics and can be solved by the same
methods\citep{LL}.
It has characteristics $C_{\pm}$ and Riemann invariants $R_{\pm}$,
\begin{equation}\label{Riemann}
C_{\pm} : \left(\frac{d\xi}{dy}\right)_{\pm}=\pm c~+u,
\qquad R_{\pm}\equiv
 u\pm\frac{2c}{\gamma-1}.
\end{equation}
The $C_+$ characteristics propagate toward the planet from the unperturbed
disk, where $u=0$ and $c=c_0$.
Therefore $R_+=2c_0/(\gamma-1)$ \emph{everywhere}, and
$u=2(c_0-c)/(\gamma+1)$.\footnote{We take $x<0$ and $y>0$, which
describes the wake interior to the orbit of the planet.  The discussion
that follows is equally valid for $x>0$ and $y<0$ if one interchanges
the roles of $C_+$ and $C_-$}
Eqs.~(\ref{nonlin1}) then reduce to a single first-order equation,
\begin{eqnarray}\label{Burger}
0&=&\partial_y \psi - (1+\psi)c_0 \partial_\xi \psi,\\[10pt]
\qquad \psi(y,\xi)&\equiv& \frac{\gamma+1}{\gamma-1}\,\frac{c-c_0}{c_0}
~~\approx\frac{\gamma+1}{2}\,\frac{\Sigma-\Sigma_0}{\Sigma_0}\quad\mbox{if}~
\psi\ll 1,\nonumber
\end{eqnarray}
which is the well-known inviscid Burger's equation: the simplest
nonlinear equation capable of displaying a shock \citep{Whitham}.

Unfortunately, the analysis above is oversimple.  In the limit of
an infinitesimal disturbance, eqs.~(\ref{nonlin1}) predict
that the wave propagates at constant amplitude.  But a linear WKB treatment
of eqs.~(\ref{nonlin0})
shows that  the amplitude of such a wave grows in proportion to
$|x|^{1/2}\propto|\xi|^{1/4}$ as a consequence of conservation of
angular momentum flux.  To recover this behavior, the larger
subdominant terms involving $v$ that were omitted in passing from
eqs.~(\ref{nonlin0}) to (\ref{nonlin1}) must be reinstated.
The equations for the $C_\pm$ characteristics become
\begin{displaymath}
\partial_{\pm} R_{\pm} \approx \frac{1}{2Ax}\left(2\Omega v \mp
c\partial_y v\right)\equiv S_\pm.
\end{displaymath}
where $R_{\pm}$ have the same meaning as before, and we have introduced
the abbreviations $\partial_\pm\equiv\partial_y+(u\pm c)\partial_\xi$
for the derivatives along $C_\pm$.  The source terms $S_\pm$
are smaller than individual terms in $R_\pm$ 
by $O(\xi^{-1})$ but nonzero, so that the $R_\pm$ are not invariant.
Yet $R_+$ is still well approximated by its undisturbed value
$2c_0/(\gamma-1)$:  First of all
$R_+$ is truly undisturbed until the incoming $C_+$ characteristic 
encounters the leading edge of the wake, which propagates in
the opposite direction on a bundle of $C_-$ characteristics of roughly
constant width $\Delta\xi$.
Then, since the encounter lasts a ``time'' $\Delta y\sim\Delta\xi /2c$, the
maximum change in $R_+$ relative to its undisturbed value is
$O(\Delta\xi/\xi)\ll 1$, which may be neglected at large $\xi$.

The source term $S_-$ cannot be neglected because it ``travels with''
the wake so that its effects accumulate.
We solve for $S_-$ using the second of
eqs.~(\ref{nonlin0}), which, to an adequate approximation, reads
\begin{displaymath}
\partial_y v\approx -\frac{1}{2Ax}\left(2Bu~+~\frac{c^2}{\Sigma}
\partial_y\Sigma\right).
\end{displaymath}
Only those parts of $S_-$ that are in phase with $u$ or
$\Sigma-\Sigma_0$ are important for the growth of the amplitude;
the out-of-phase parts make a slight but negligible change
to the characteristic velocity.  The operator $\partial_y$ changes
the phase by $90^\circ$, so from the  equation above,
\begin{displaymath}
-\frac{1}{2Ax}\left[c\partial_y v +2\Omega v\right]_{\rm in~phase}=
-\frac{1}{2A\xi} \left(2Bcu~+~2\Omega\frac{c^2-c_0^2}{\gamma-1}\right)
\end{displaymath}
We approximate $c$ by $c_0$ except in the difference $c-c_0$ and
invoke the approximate constancy of $R_+$  to write
$u\approx 2(c_0-c)/(\gamma-1)$.
Finally we write $\psi\equiv(\gamma+1)(c-c_0)/(\gamma-1)c_0$ as before,
and so obtain the corrected $C_-$ equation in the form
\begin{equation}\label{sourced}
\partial_y\psi-(1+\psi)c_0\partial_\xi\psi = -\frac{c_0}{4\xi}\psi.
\end{equation}

As expected, the general solution of the \emph{linearized} form of
eq.~(\ref{sourced}) is $\psi=|\xi|^{1/4}f(\xi+c_0y)$, 
where $f$ is an arbitrary function.
The full equation incorporates all three processes
that dominate propagation of the wake:\\
(i) increasing radial wavenumber with distance from the planet
($\partial_x=2Ax\partial_\xi$);\\
(ii) increasing amplitude due to conservation of angular-momentum flux;\\
(iii) nonlinear steepening.

Fortunately, eq.~(\ref{sourced}) can be transformed
back into Burger's equation by changing the independent and dependent
variables.
First let $\xi\to c_0l(\eta\,-y/l)$ (so $\eta$ is dimensionless):
\begin{displaymath}
l\partial_y\psi-\psi\partial_\eta\psi=\frac{l}{4y}\psi.
\end{displaymath}
Since the wake is confined to $|\eta|\ll y/l$,
we have approximated $y-\eta l$ by $y$ on the righthand side.
Next, let $\psi=(y/l)^{1/4}\chi$ to absorb the source term:
\begin{displaymath}
l\partial_y\chi-(y/l)^{1/4}\chi\partial_\eta\chi=0.
\end{displaymath}
Finally, introduce a new dimensionless ``time'' (or azimuth)
variable ${t}$ as in eq.~(\ref{taudef}), so that the dimensionless
equation of motion becomes (\ref{chieqn}).

\end{document}